\documentclass[twocolumn,prb,showpacs,superscriptaddress]{revtex4}
\usepackage[dvips]{graphicx}%
\usepackage{dcolumn}
\usepackage{amsmath}
\begin{document}

\title{Direct measurement of molecular stiffness and damping in confined water layers}
\author{Steve Jeffery}
\altaffiliation{Current address: Oak Hill Theological College,
Southgate, London N14 4PS , UK} \affiliation{Department of
Materials, University of Oxford, Parks Road, Oxford OX1 3PH,
United Kingdom}
\author{Peter M. Hoffmann}
\email{hoffmann@physics.wayne.edu} \affiliation{Department of
Physics, Wayne State University, 666 W. Hancock, Detroit MI
48201, USA}
\author{John B. Pethica}
\altaffiliation{Current address: Department of Physics, Trinity
College, Dublin, Ireland} \affiliation{Department of Materials,
University of Oxford, Parks Road, Oxford OX1 3PH, United Kingdom}
\author{Chandra Ramanujan}
\affiliation{Department of Materials, University of Oxford, Parks
Road, Oxford OX1 3PH, United Kingdom}
\author{H. \"Ozg\"ur \"Ozer}
\affiliation{Department of Physics, Trinity College, Dublin,
Ireland}
\author{Ahmet Oral}
\affiliation{Department of Physics, Bilkent University, Ankara, Turkey}
\date{\today}

\begin{abstract}
We present {\em direct} and {\em linear} measurements of the
normal stiffness and damping of a confined, few molecule thick
water layer. The measurements were obtained by use of a small
amplitude (0.36 $\textrm{\AA}$), off-resonance Atomic Force
Microscopy (AFM) technique. We measured stiffness and damping
oscillations revealing up to 7 layers separated by 2.56 $\pm$ 0.20
$\textrm{\AA}$. Relaxation times could also be calculated and were
found to indicate a significant slow-down of the dynamics of the
system as the confining separation was reduced. We found that the
dynamics of the system is determined not only by the interfacial
pressure, but more significantly by solvation effects which depend
on the exact separation of tip and surface. Thus \lq
solidification\rq\, seems to not be merely a result of pressure
and confinement, but depends strongly on how commensurate the
confining cavity is with the molecule size. We were able to model
the results by starting from the simple assumption that the
relaxation time depends linearly on the film stiffness.
\end{abstract}
 \maketitle

\section{Introduction} The structure of water, as the primary
biological solvent, has been intensively studied. For example,
liquid water adopts short-range order, which depends strongly on
dissolved species or geometric constraints. This structure emerges
from the minimization of the free energy associated with the
dynamic system of hydrogen bonds between neighboring water
molecules. The entropy cost of the induced order almost certainly
plays an important role in determining the structure of biological
molecules that depend on hydration for their function, such as
proteins and cell membranes\cite{finney96}.

A surface can act as a model system for studying these phenomena,
as it is both geometrically disrupting and can be chemically
functionalized to affect the structure of the water close to it. A
particularly interesting problem is the emergence of density
oscillations as a function of film thickness when water is
confined between two surfaces\cite{israelachvili}. This phenomenon
is related to the radial density fluctuations in solvation shells
of solutes. These density fluctuations have been originally
observed by diffraction methods in clay-water
systems\cite{bradley37, delpennino81}.

In 1982, the mechanical response of confined water layers was
directly determined using the Surface Force Apparatus
(SFA)\cite{israelachvili83}. With the invention of the Atomic
Force Microscopy (AFM), attempts were made to measure stiffness
oscillations with this technique\cite{oshea92}. AFM probes have a
much smaller contact area than SFA.  This is an advantage if local
changes in the water structure are to be examined\cite{patrick97},
and potentially allows for probing regions of negative contact
stiffness. The latter cannot be probed using SFA, because the
instrument stiffness is not high enough to withstand the snap-in
instability in negative stiffness regions.  The disadvantage of
AFM is that the signals are much smaller and the contact area is
determined by the tip shape and thus is essentially unknown.  The
small signal-to-noise ratio in AFM made the direct measurement of
water structure an elusive goal. In 1995, Cleveland et al.\
\cite{cleveland95} measured the oscillatory potential of the
confined water layers indirectly by analyzing the Brownian noise
spectrum of a AFM tip immersed in water.  More recently, direct
measurements of the structure were achieved by Jarvis et al.\
\cite{jarvis00} by using nanotube probes and a large amplitude AFM
technique, and by Antognozzi et al.\ \cite{antognozzi01} who
measured the local shear modulus using an AFM in shear force mode.

In this paper we present results of direct and {\em linear}
measurements of the normal junction stiffness of water confined
between the AFM tip and an atomically smooth mica surface.  This
was achieved by using ultra-small amplitudes of 0.36
$\textrm{\AA}$ and sub-resonance operation, which avoids the
problem of reduced quality factor in liquids.  The snap-in
instability was avoided by using a sufficiently stiff cantilever
(here 0.65 N/m). This method is ideal to make quantitative,
point-by-point measurements of the mechanical properties of
confined water layers.  The small amplitudes (much smaller than
the nominal size of a water molecule) allow us to measure the
elastic and viscous response of the confined water layer without
disrupting the layers themselves, as would be the case in the
large amplitude methods used previously. The challenge of such a
technique is the measurement of exceedingly small signals, since
the usual methods of signal enhancement (large amplitudes, low
stiffness levers, resonance operation) are not used.  Recently, we
succeeded in implementing such a technique in UHV\cite{oral03,
hoffmann01a, oral01} -  and in liquids\cite{hoffmann01b}, using an
improved fiber interferometric displacement sensor\cite{rugar89}
to overcome the reduced signal-to-noise of the technique. Here we
report on our direct measurements of the mechanical properties of
confined water layers using this novel AFM technique.

\section{Experimental} Small amplitude, off-resonance AFM\cite{oral03} has
recently been successfully used for measuring atomic bonding
curves\cite{hoffmann01a}, mapping force gradients at atomic
resolution\cite{oral01}, and measuring atomic scale energy
dissipation\cite{hoffmann01c}. Both the force gradient and the
damping coefficient/ energy dissipation can be obtained by solving
the equation of motion for a forced damped oscillator at a drive
amplitude $A_0 \ll \lambda$ (where $\lambda$ is the nominal range
of the measured interaction) and $\omega \ll \omega_0$.  The
equation of motion is given by:
\begin{equation}
m \ddot{x}+\gamma \dot{x}+(k_L+k)x=k_L A_0 \exp(i\omega t)
\end{equation}
where we linearized the force field, owing to the fact that lever
amplitudes are much smaller than the range of the measured
interactions, $\lambda$. This assumption has recently been shown
to be justified if $A_0$ is sufficiently small\cite{hoffmann03}.
After solving the equation, we find for the interaction stiffness
and the damping coefficient:
\begin{equation}\label{stiffness}
k=k_L (\frac{A_0}{A}\cos\phi-1)
\end{equation}
and
\begin{equation}\label{damping}
\gamma=-\frac{k_L A_0}{A \omega}\sin\phi
\end{equation}
Here, $A_0$ is the drive amplitude of the lever, $A$ is the
measured amplitude as the surface is approached, $k_L$ is the
lever stiffness, $\phi$ is the measured cantilever phase, and $k$
is the measured interaction stiffness.  In equation
$\eqref{damping}$, $\omega$ is the drive frequency and $\gamma$ is
the damping coefficient. In performing the above calculations, we
intrinsically assume that elastic and viscous forces are additive,
i.\ e.\ they act in parallel (Kelvin model). In modeling liquids,
however, typically a Maxwell model is used in which the elastic
and viscous (damping) term are considered to be in series. To
convert from one to the other we can use the following set of
equations\cite{findley}:
\begin{eqnarray}\label{MKdamping}
\eta=\gamma+\frac{k^2}{\omega^2 \gamma} \\
R=k+\frac{\omega^2 \gamma^2}{k}\label{MKstiffness}
\end{eqnarray}
where $\eta$ and $R$ are the viscous and elastic terms in the
Maxwell model, respectively.

The used AFM was home-built and incorporated a fiber
interferometer mounted on a remote controlled nano-manipulator
with 5 degrees of freedom and a step size of $<$ 100 nm. The
nominal sensitivity of the interferometer was $3\times 10^{-4}
\textrm{\AA}/\sqrt{\textrm{Hz}}$. This allowed us to measure
stiffnesses of a few $10^{-2}$ N/m using a 0.65 N/m cantilever, a
sub-Angstrom lever amplitude of 0.36 $\textrm{\AA}$ and reasonable
integration times. The measurement frequency was 411 Hz.
Measurements were performed in ultrapure water with a
concentration of 0.01 M KCl. The surface was freshly cleaved mica,
and the cantilever tip was made out of silicon.
\\
\section{Results}

\begin{figure}\centerline{\includegraphics[width=83mm, clip,
keepaspectratio]{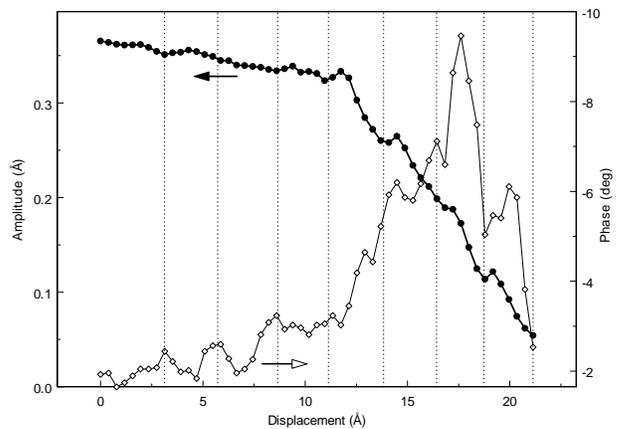}} \caption{Amplitude and phase measured on a
water layer confined between the AFM tip and a mica surface. The
mica surface is located to the right. Several oscillations can be
seen in the amplitude data. The overall decrease as the surface is
approached is due to hydrophilic effects. The phase shows a more
complicated behavior (discussed in text), but also shows clear
oscillations. The reference lines correspond to displacements
where liquid ordering is maximized and serve as a guide for the
eye.}\label{ampphase}
\end{figure}

Figure \ref{ampphase} shows the amplitude and the cantilever phase
as a function of displacement. The surface is located to the right
of the graph, and the monotonic drop-off of the amplitude as the
surface is approached can be attributed to repulsive interactions,
which are most likely hydrophilic in origin. The amplitude data
shows at least 5 equally spaced local minima (and maxima). The
phase data shows equally spaced maxima further away from the
surface, which roughly line up with the minima of the amplitude
data. However, as the surface is further approached additional \lq
intermediate\rq\, peaks appear close to the amplitude maxima, and
these peaks finally dominate as the gap is decreased to a few
molecular spacings.  The average spacing between the amplitude
minima (and the phase maxima further out) is $2.56 \textrm{\AA}
\pm 0.20 \textrm{\AA}$, consistent with earlier
reports\cite{israelachvili83, patrick97, cleveland95, jarvis00,
antognozzi01}. Overall, the phase increases up to a global maximum
as the surface is approached and then decreases again closer to
the surface.

\begin{figure}\centerline{\includegraphics[width=83mm, clip,
keepaspectratio]{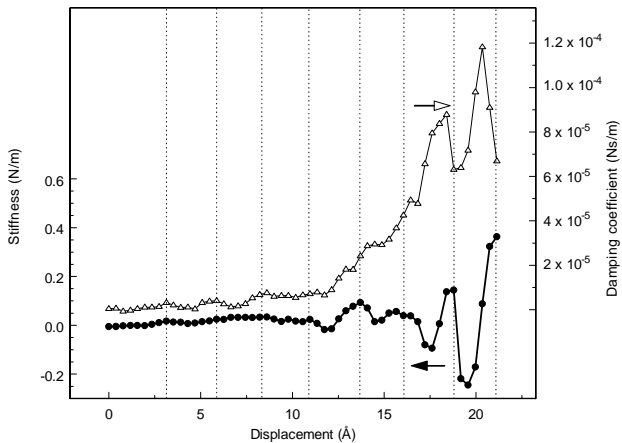}} \caption{Solvation stiffness and (Kelvin)
damping coefficient versus displacement. The solvation stiffness
was obtained by calculating the stiffness from equation
$\eqref{stiffness}$ and subtracting the exponential background.
Again, clear oscillations spaced at 2.56 $\textrm{\AA}$ can be
seen in both the stiffness and the damping. The damping exhibits a
\lq phase-shift\rq\, with respect to the stiffness data at a
displacement of 15 $\textrm{\AA}$. Closer to the surface the
damping is out-of-phase with the stiffness, while further away it
switches to being in-phase.}\label{stiffdamp}
\end{figure}

The measured stiffness (equation $\eqref{stiffness}$) can be
decomposed into two components: A monotonic background, and an
oscillatory term, which is the one we will be most concerned with
in this paper. The monotonic background is most likely due to
double-layer (DVLO) and hydrophilic interactions, which can both
be modelled as exponentials.  A best fit and subsequent
subtraction of the monotonic background yields the data shown in
Figure \ref{stiffdamp}. Also shown in Figure \ref{stiffdamp} is
the damping coefficient calculated using equation
$\eqref{damping}$. Note that the peaks in the stiffness data
correspond to the minima in the amplitude data and thus to the
higher stiffness of the ordered phase of the confined water layer.
Close to the surface, the damping curve shows peaks that are \lq
out of phase\rq\, with the stiffness data. Further away from the
surface, however, double peaks occur, and finally the damping
shows peaks that are in-phase with the stiffness maxima, similar
to the phase data shown in Figure \ref{ampphase}. Due to the
dissipation, the cantilever loses kinetic energy. The energy loss
per cycle can be calculated from\cite{anczykowski99}
\begin{equation}
E_{\textrm{diss}}=\oint_{\textrm{cycle}}\gamma
\dot{x}=\pi\gamma\omega A^2=\pi k_L A_0 A \sin \phi
\end{equation}
The maximum loss was $E_\textrm{diss}$ = 1.3 meV per cycle, which
was observed close to the maximum in the phase (Figure
\ref{ampphase}).

What is the origin of the observed dissipation and its increase as
the surface is approached? One interpretation would be to
attribute the damping to viscous drag, especially due to the
squeezing of the liquid between the tip and the
substrate\cite{oshea98}. The squeeze Reynolds number, Re, is given
by:
\begin{equation}
\textrm{Re}=\frac{\rho_W z}{\eta_W}\frac{\textrm{d}z}{\textrm{d}t}
\end{equation}
where $\eta_W$ is the viscosity of water, $\rho_W$ is its density,
and $z$ is the tip-surface distance. In our case, $z \approx 1$
nm, $\textrm{d}z/\textrm{d}t \approx \omega  A \approx 100$ nm/s
and we obtain Re $\approx 10^{-10} \ll 1$. In the following we
assume that the viscous damping due to the cantilever beam does
not change much with separation (since the beam is several
micrometers away), and thus the variation of the damping with
separation is dominated by the damping at the tip. With Re$\ll$1,
the squeeze damping term between tip and surface is given
by\cite{oshea98}:
\begin{equation}
\gamma_s=6\pi\eta_W\frac{R_\textrm{tip}^2}{z}
\end{equation}
Using reasonable values (10-100 nm) for the tip radius,
$R_\textrm{tip}$, we find that the expected viscous damping at
less than 1 nm separation is of the order of $10^{-9}$ to
$10^{-7}$ Ns/m, which is about three to four orders of magnitude
smaller than the measured values (see Figure \ref{stiffdamp}),
which are of order $10^{-5}$ to $10^{-4}$ Ns/m. We found that the
viscosity increased exponentially with distance and thus is large
only very close to the surface ($<$ 1 nm). This effect has been
observed before and has been attributed to a sharp increase in the
effective viscosity of confined liquid layers\cite{antognozzi01,
oshea98, zhu01}, a possible indication of the altered dynamical
and structural properties of liquids under confinement. However,
Raviv et al.\ \cite{raviv01, raviv02} recently observed
viscosities close to the bulk value even at separations as small
as 1 nm. Their results were based on measurements of the snap-in
instability close to the surface, while in the present experiment
the mechanical properties of the film were measured continuously
without any mechanical instabilities. Moreover, in our experiment
we considered normal forces, while their results are based on
shear measurements. How these measurements relate to each other
and if there is a fundamental difference between the normal and
the lateral dynamic behavior of water is an important question for
future study.

\begin{figure}\centerline{\includegraphics[width=83mm, clip,
keepaspectratio]{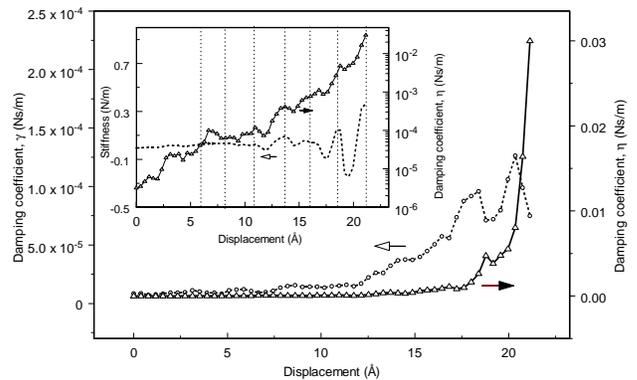}} \caption{Comparison of Kelvin damping,
$\gamma$, and Maxwell damping, $\eta$.  If we treat the confined
film as a liquid (Maxwell-type model), we find that the damping
increases very strongly as the film is squeezed to a few molecular
layers, unlike the Kelvin damping which increases more moderately.
The inset shows the Maxwell damping, $\eta$, on a log- scale, with
the stiffness as a reference. It can be seen that $\eta$ is
essentially in-phase with the stiffness throughout the measurement
range.}\label{maxwell}
\end{figure}

Another debate has been the nature of the structural and dynamical
differences between confined and bulk liquids. The difference has
been attributed to a type of first-order phase transformation from
liquid to solid\cite{klein98}, or, alternatively, to a continuous
transformation, not unlike a glass transition\cite{demirel96}. If
the liquid does indeed turn solid under certain confinement
conditions, the proper mechanical model would be a Kelvin-type
model. However, if the liquid stays essentially liquid albeit with
greatly enhanced viscosity (recent evidence for this comes from
diffusion measurements\cite{mukhopadhyay02}), a Maxwell-type model
should be used. As in any \lq standard\rq\, analysis of AFM, we
used a Kelvin-type model above. However, to elucidate the nature
of the changes under confinement further, it is important to use a
model that properly applies to liquids. Using equations
$\eqref{MKdamping}$ and $\eqref{MKstiffness}$ we transformed the
measured stiffness and damping terms to the Maxwell model.  We
found that the stiffness remains almost unchanged between the two
models (i.\ e.\ $k \approx R$), but as shown in Figure
\ref{maxwell}, the viscous term changes dramatically. The Kelvin
model damping term, $\gamma$, is out-of-phase with the stiffness
oscillations close to the surface (Figure \ref{stiffdamp}), while
the Maxwell model damping, $\eta$, is much larger and essentially
in-phase with the stiffness variations (Figure \ref{maxwell}). As
mentioned above, further away from the surface, the Kelvin damping
experiences a \lq phase shift\rq\, and becomes in-phase with the
stiffness (similar to the phase data in Figure \ref{ampphase}).
The Maxwell damping, on the other hand, remains in-phase
through-out. More about this below.

\begin{figure}\centerline{\includegraphics[width=83mm, clip,
keepaspectratio]{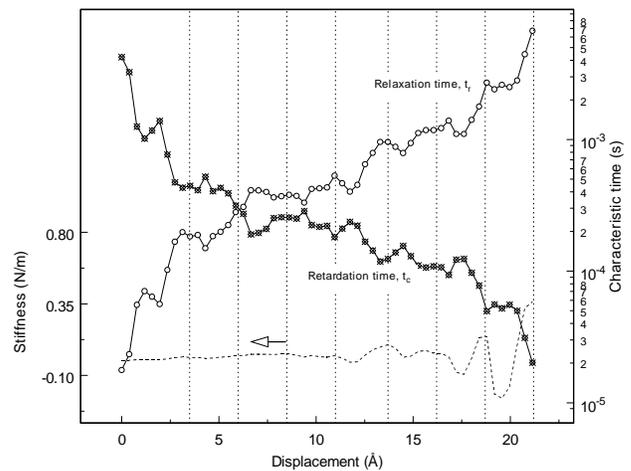}} \caption{Relaxation ($t_r$) and
retardation ($t_c$) time plotted versus displacement. The
relaxation time is in-phase with the stiffness, while the
retardation time is out-of-phase. The relaxation time increases
overall due to the increasing pressure, but also exhibits in-phase
oscillations due to solvation effects.}\label{relax}
\end{figure}

When dealing with dissipative behavior it is useful to look at the
characteristic time constants involved in the dynamic behavior of
the confined liquid. In the Kelvin model, a characteristic time is
given by $t_c=\gamma /k$, which is called the \lq retardation
time\rq\, \cite{findley}. This time is approximately the time
needed to build up a significant strain in the material upon
application of a constant stress. In standard solids, a certain
amount of strain can be obtained almost instantaneously due to the
elasticity of the material, however, in ideal liquids,
instantaneous strain is not possible due to the velocity-dependent
damping. Thus a lower $t_c$ might indicate a more solid-like
material. On the other hand, in the Maxwell model, the
characteristic time is $t_r=\eta /R$, which is the \lq relaxation
time\rq\,. This time is related to the time needed for stresses in
the material to relax after a strain has been imposed. In solids,
stresses will persist for long times when a strain is applied (one
of the characteristic of materials being solid), while in liquids
any stresses will quickly dissipate away. Thus higher $t_r$
indicates a more solid-like behavior. It should be noted that
$t_c$ and $t_r$ are simply related by:
\begin{equation}\label{times}
t_c=\frac{1}{\omega^2}{t_r}
\end{equation}
The dependence of $t_c$ and $t_r$ on displacement is shown in
Figure \ref{relax}. It can be seen that overall $t_c$ is
decreasing and $t_r$ increasing as the liquid layer is
increasingly confined. This indicates a tendency for the layer to
become more solid-like. Even more interesting, however, is the
fact that at separations where the stiffness oscillations are at
their maximum, $t_c$ is lowered and $t_r$ is increased. This is a
further indication that the water becomes more solid-like when it
is allowed to order, i.\ e.\ when the tip-surface separation is
commensurate with the \lq natural\rq\, molecular spacing of water.
In the \lq ordered phase\rq\,, the stiffness $k$ (or $R$) is
maximum, the retardation time, $t_c$, is minimum, and the
relaxation time, $t_r$, is maximum, as expected for a solid.

Another interesting observation is that the characteristic times
$t_r$ and $t_c$ change very slowly with separation until a few
Angstrom from the surface. Then they seem to change more rapidly,
with the liquid becoming seemingly even more solid-like. Some
authors\cite{klein98} have argued that this behavior shows the
liquid undergoes some kind of first order phase transformation
upon confinement, while others suggest a more gradual, glass-like
transition\cite{demirel96}. Recent results measured under shear
even suggest that water in particular fails to \lq solidify\rq\,
at all\cite{zhu01}. Based on our results we cannot conclusively
decide between these viewpoints. However, as we will see below,
there seems to be both a gradual stiffening of the layer as
pressure is applied and a much more pronounced
periodic change of the mechanical behavior of the layer.\\

\section{Discussion and Model}

There are several surprising findings from the linear measurement
of confined water presented here: 1) Observed oscillations in the
phase and dissipation extend much further than the oscillations in
the amplitude or stiffness, 2) the phase and the Kelvin damping
oscillations experience a \lq phase shift\rq\, with respect to the
stiffness data as we move away from the surface, 3) while on
average the amplitude continuously decreases as the surface is
approached, the phase seems to pass through an intermediate
maximum, and, finally, as hinted above, 4) the mechanical behavior
of the layer changes both gradually (hydrophilic background) and
more abruptly (solvation shell oscillations). To explain this
behavior we simulated the nanomechanical behavior of the water
layer by starting from the assumption that the relaxation time,
$t_r$, is to first order linearly dependent on the stiffness of
the water layer. This is not to be taken literally, in the sense
of a direct physical connection between the stiffness and the
relaxation time (although there well might be), but rather the
stiffness is seen as an indicator of the \lq solidness\rq\, of the
layer, and the relaxation time (as another indicator) is taken to
be essentially proportional to it. We found that we can get the
best fit of our data if we assume that the relaxation time depends
linearly on both the background stiffness, $k_h$, due to
hydrophilic interaction and on the stiffness oscillations, $k_s$,
due to the solvation effects but with two different \lq
coupling\rq\, constants $\alpha_1$ and $\alpha_2$:
\begin{equation}
t_r=\alpha_1 k_h+\alpha_2 k_s +t_0
\end{equation}
Here, $t_0$ corresponds to the relaxation time measured far away
from the surface. The advantage of using separate constants
$\alpha_1$ and $\alpha_2$ is that we can separate the effect of
background hydrophilic interactions from the influence of
solvation forces on the relaxation time. We found that in order to
reproduce the experimental results as closely as possible it was
necessary to set $\alpha_1$ to $2.3 \times 10^{-3}$ sm/N and
$\alpha_2$ to $5 \times 10^{-3}$ sm/N, i.\ e.\ the relaxation time
was more than twice as sensitive to solvation forces than it was
to the hydrophilic background. It should be noted that the
hydrophilic background stiffness is directly proportional to the
load (or surface pressure), since both are exponentials and one is
the derivative of the other. Thus by the above approach we can
separate the effects of the overall pressure or load from the
effect of the liquid ordering which only occurs at certain,
molecularly commensurate separations.

The retardation time, $t_c$, can be calculated from equation
$\eqref{times}$. The damping coefficients $\gamma$ and $\eta$ are
then given by:
\begin{eqnarray}\label{gamma}
\gamma=t_c\cdot k \\
\eta=t_r\cdot R
\end{eqnarray}
In the simulations, we took $k = R$ (as found experimentally) to
simplify the calculations.  The solvation force was modeled as
follows\cite{israelachvili}:
\begin{multline}\label{fs}
F_s=\sum_{\textrm{tip},k=0}^{N} 2\pi r ^2(z_k) \cdot \\ \cdot k_B
T \rho \cos \left(\frac{2\pi (z_k+D)}{\sigma}\right)\exp
\left(-\frac{z_k+D}{\sigma}\right)
\end{multline}
where we summed the contributions of different areas of the tip by
subdividing the tip into $N$ horizontal \lq slices\rq\, of radius
$r(z_k)$. Here, $\rho$ is the particle density of water, $\sigma$
is both the period and the decay parameter of the interaction
(they were experimentally found to be nearly identical), $z_k$ is
the height of the $k$'th slice of the tip, and $D$ is the
tip-surface separation. The hydrophilic interaction is given by:
\begin{equation}\label{fh}
F_h=\sum_{\textrm{tip}, k=0}^{N} 2\pi r^2(z_k)\cdot p_h
\exp\left(-\frac{z_k+D}{\lambda}\right)
\end{equation}
where $p_h$ is a constant and $\lambda$ is the decay parameter of
the hydrophilic background.  All parameters in expressions
$\eqref{fs}$ and $\eqref{fh}$ were determined from the experiment.
The hydrophilic decay parameter was found to be only slightly
smaller ($\lambda = 2.45 \textrm{\AA}$) than the decay parameter
of the oscillations ($\sigma = 2.56 \textrm{\AA}$). The
corresponding stiffnesses were found from taking the derivative of
the forces with respect to tip-surface separation, $k =
-\textrm{d}F/\textrm{d}D$.

\begin{figure}\centerline{\includegraphics[width=83mm, clip,
keepaspectratio]{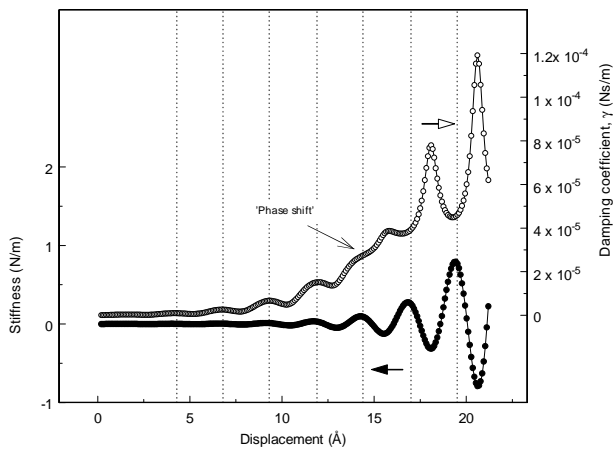}} \caption{Simulated stiffness and Kelvin
damping, $\gamma$. Compare to measured data (Figure
\ref{stiffdamp}). Although we had to assume larger stiffness
oscillations in the model, the overall agreement is good, and the
\lq phase shift\rq\, in the damping data is reproduced
well.}\label{kelvinsim}
\end{figure}

Since we cannot know the exact geometry of the tip, we did not
expect to get a perfect agreement between theory and experiment.
Nevertheless we obtained a semi-quantitative agreement that
reproduces all of the surprising features mentioned above. The
geometry of the tip was assumed to be paraboloid, and the best
agreement with experimental data was obtained for a nominal radius
of 1nm. Figure \ref{kelvinsim} shows the calculated total
stiffness and Kelvin damping coefficient, $\gamma$. The damping
coefficient is out-of-phase with the stiffness close to the
surface, then undergoes a \lq phase shift \rq\, and becomes
in-phase close to the surface, as seen in the experiment. From
equation $\eqref{gamma}$ we see that the damping is a product of
the retardation time, $t_c$ and the stiffness, $k$. Close to the
surface the damping is dominated by the retardation time $t_c$,
which is always out-of-phase with the stiffness (Figure
\ref{relax}), while further away from the surface the stiffness
$k$ dominates the variation in the damping. On the other hand, the
Maxwell damping, $\eta$, is always in phase with the stiffness,
since the relaxation time $t_r$ is in-phase with $R$ (or $k$).

It can be shown that a more complicated mechanical model, such as
the commonly used Burger's model, behaves like a Maxwell model at
low frequencies. This implies that the present discussion has more
general implications than might be expected from the use of such
simplified mechanical models. In particular, it would seem from
equation $\eqref{gamma}$ that the oscillatory behavior of the
Kelvin damping, $\gamma$, could be explained by the oscillatory
behavior of the stiffness, even if the retardation time is
constant or slowly varying. In this scenario, the observed
oscillations of the retardation/relaxation time would be merely an
\lq artifact\rq\, of the calculation. However, if the retardation
time were constant or smoothly varying (i.e. not oscillating), the
Kelvin damping would have to remain in-phase with the stiffness at
all times. This is not observed in the experiment. The fact that
$\gamma$ is out-of-phase close to the surface implies that the
relaxation/retardation time of the liquid exhibits
separation-dependent oscillations independent of the oscillations
of the stiffness. This means that the {\em dynamic} behavior of
the confined liquid is strongly affected by how commensurate the
tip-surface spacing is with regard to the size of the confined
molecules.

\begin{figure}\centerline{\includegraphics[width=83mm, clip,
keepaspectratio]{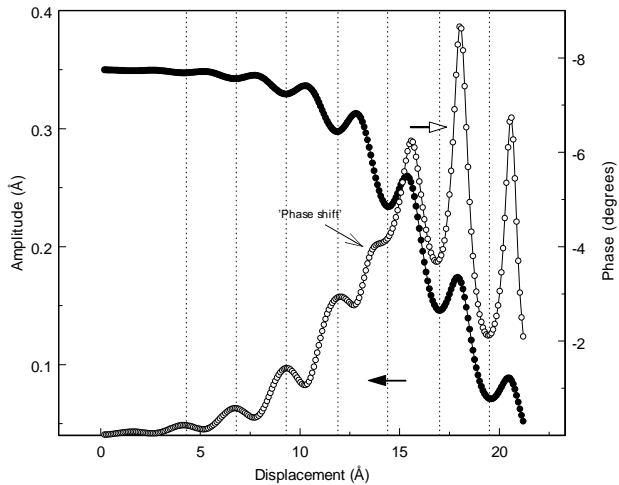}} \caption{Simulated amplitude and phase.
Compare to measured data (Figure \ref{ampphase}). There is good
qualitative agreement and the complicated behavior of the phase is
well reproduced including the \lq phase shift\rq\, with respect to
the stiffness (here: amplitude) and the global
maximum.}\label{phase}
\end{figure}

The phase was calculated by solving equations $\eqref{stiffness}$
and $\eqref{damping}$ simultaneously (and assuming $\omega \ll
\omega_0$):
\begin{equation}\label{tanphi}
\tan\phi=\frac{\omega\gamma}{k+k_L}
\end{equation}
The simulated phase, $\phi$, is shown in Figure \ref{phase}. The
simulation reproduces all the \lq puzzling\rq\, features of the
experiment: The shift from being out-of-phase to being in-phase
with the stiffness oscillations and the intermediate maximum in
the phase. The shift is due to the shift in $\gamma$ discussed
above. The intermediate maximum is due to the fact that the
stiffness changes slowly far from the surface, but then rather
rapidly closer in, \lq overtaking\rq\, the damping coefficient in
the process (equation $\eqref{tanphi}$). The observation that
oscillations in the phase or damping are observable further away
from the surface than the oscillations of the stiffness can also
be explained: As we can see in the simulated phase, a phase of
more than $1^\circ$ is observed as far away as $13 \textrm{\AA}$
from the closest approach (about 5 water layers). Such a phase
angle can be easily measured with a lock-in amplifier. On the
other hand, at the same separation, the stiffness is only 0.04 N/m
requiring a measurement of a change in lever amplitude of the
order of $0.02 \textrm{\AA}$, which is more difficult to measure
and can be lost in the noise.

In conclusion, we can see that our simple approach of directly
relating the relaxation time to the stiffness of the layer has
allowed us to reproduce all the important features of the
experiment. The relaxation/retardation times can therefore be
taken as fundamental physical parameters (together with the
stiffness) that characterize the mechanical properties of the
system quite well. The weaker dependence of the relaxation time on
the hydrophilic interaction and the more pronounced dependence on
solvation forces, suggests a compromise in the continuing debate
over the nature of the solid-liquid transition. It seems that
there is a gradual increase of the relaxation time with surface
pressure and a more substantial change related to the molecular
ordering of the liquid close to the surface.

\end{document}